\RequirePackage{amsmath}
\documentclass[runningheads]{llncs}

\usepackage{cite}
\usepackage{bbding}
\usepackage{graphicx}
\usepackage{comment}
\usepackage{url}
\usepackage[utf8]{inputenc}
\usepackage{amsmath}
\usepackage{amsfonts}
\usepackage{amssymb}
\usepackage{listings}
\usepackage{bussproofs}
\usepackage{float}
\usepackage{xspace}
\usepackage{xifthen}
\usepackage[nomap]{FiraMono}
\usepackage{tikz}
\usepackage{subcaption}
\usepackage{framed}
\usepackage{stmaryrd} 
\usetikzlibrary{positioning}
\usepackage[ruled,vlined,linesnumbered]{algorithm2e}
    \SetAlFnt{\fontsize{8.5}{10}\selectfont}
    \SetArgSty{textup}
    \SetProgSty{textup}
    \SetFuncArgSty{textup}
    \SetFuncSty{textit}
    \SetKwProg{procedure}{def}{}{}
    \SetKwProg{datastructure}{Datastructure}{}{}
    \SetKwSwitch{Switch}{Case}{Other}{match}{:}{case}{otherwise}{endcase}{endsw}
    
    \SetKw{KwCont}{continue}
    \SetKw{KwBreak}{break}

\SetKwComment{Comment}{/* }{ */}

\makeatletter
\RequirePackage[bookmarks,unicode,colorlinks=true]{hyperref}%
   \def\@citecolor{blue}%
   \def\@urlcolor{blue}%
   \def\@linkcolor{blue}%

\def\orcidID#1{\smash{\href{http://orcid.org/#1}{\protect\raisebox{-1.25pt}{\protect\includegraphics{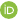}}}}}
\makeatother

\spnewtheorem{thm}{Theorem}{\bfseries}{\itshape}
\spnewtheorem{prop}{Proposition}{\bfseries}{}
\spnewtheorem{cor}{Corollary}{\bfseries}{\itshape}
\spnewtheorem{defin}{Definition}{\bfseries}{}
\spnewtheorem{ex}{Example}{\bfseries}{}

\definecolor{green}{rgb}{0, 0.6, 0}

\definecolor{comments}{RGB}{80,0,110}

\lstdefinelanguage{scala}{
    alsoletter={@,=,>},
    keywordstyle = {\color{black}\bfseries},
    keywordstyle = [2]{\color{black}\bfseries},
    commentstyle = \color{comments},
    morekeywords = [2]{abstract, case, class, def, do, Input, Output, then,
        else, extends, false, free, if, implicit, match,
        object, true, val, var, while, sealed, or,
        for, dependent, null, type, with, try, catch, finally,
        import, final, return, new, override, this, trait,
        private, public, protected, package, throw},
    sensitive = true, 
    numbers=left,
    stepnumber=1,
    morecomment = [l]{//},
    morecomment = [s]{/*}{*/},
    morestring = [b]",  
    otherkeywords = {;,<<,>>,++},
    literate = {=>}{{$\Rightarrow~$}}2
}

\newcommand{\OL}{{O{\kern-0.1em}L}}
\newcommand{\POL}{{P{\kern-0.1em}O{\kern-0.1em}L}}
\newcommand{\QOL}{{Q{\kern-0.1em}O{\kern-0.1em}L}}

\newcommand{\QBF}{{Q{\kern-0.1em}B{\kern-0.1em}F}}
\newcommand{\FOL}{{F{\kern-0.1em}O{\kern-0.1em}L}}
\newcommand{\CL}{{C{\kern-0.1em}L}}

\newcommand{\FOLm}{\mathit{F{\kern-0.1em}({\kern-0.1em}O{\kern-0.1em}L{\kern-0.1em})\textsuperscript{2}}}

\newcommand{\ssimw}[1]{{{\kern-0.05em}/{\kern-0.08em}{#1}}}

\newcommand{\sime}{\sim_{{\kern-0.1em}E}}
\newcommand{\siml}{\sim_{{\kern-0.1em}L}}

\newcommand{\eqclass}[2]{{\left[#1\right]}_{#2}}
\newcommand{\eqclasssim}[2]{\eqclass{#1}{\sim_{{\kern-0.1em}#2}}}

\newcommand{\term}[2]{\mathcal{T}\ifthenelse{\isempty{#2}}%
    {}%
    {_#2}%
    \ifthenelse{\isempty{#1}}%
    {}%
    {(#1)}%
}

\newcommand{\dashvdash}{\dashv \vdash}

\newcommand{\freevars}[1]{\ensuremath{\textsf{FV}(#1)}\xspace} 

\newcommand{\OLRule}[1]{\text{#1}\xspace}
\newcommand{\ruleHypothesis}{\OLRule{Hyp}}
\newcommand{\ruleCut}{\OLRule{Cut}}
\newcommand{\ruleWeaken}{\OLRule{Weaken}}
\newcommand{\ruleLeftAnd}{\OLRule{LeftAnd}}
\newcommand{\ruleRightAnd}{\OLRule{RightAnd}}
\newcommand{\ruleLeftOr}{\OLRule{LeftOr}}
\newcommand{\ruleRightOr}{\OLRule{RightOr}}
\newcommand{\ruleLeftNot}{\OLRule{LeftNot}}
\newcommand{\ruleRightNot}{\OLRule{RightNot}}
\newcommand{\ruleLeftExists}{\OLRule{LeftExists}}
\newcommand{\ruleRightExists}{\OLRule{RightExists}}

\newcommand{\Var}{{\mathit{Var}}}
\newcommand{\interp}[2]{\left\llbracket #1 \right\rrbracket_{#2}}

\title{Interpolation and Quantifiers in Ortholattices} 
\author{
    Simon Guilloud(\Envelope)\orcidID{0000-0001-8179-7549} \and
    Sankalp Gambhir \orcidID{0000-0001-5994-1081} \and
    Viktor Kun\v{c}ak\orcidID{0000-0001-7044-9522}
}
\institute{
    EPFL \\ 
    School of Computer and Communication Sciences \\ 
    Lausanne, Switzerland \\ 
\email{\{firstname.lastname\}@epfl.ch}}

\begin{document}

\renewcommand{\vec}[1]{\overrightarrow{#1}}
\addtolength{\textfloatsep}{-0.2in}
\sloppy

\maketitle

\begin{abstract}
We study quantifiers and interpolation properties in \emph{orthologic}, a non-distributive weakening of classical logic that is sound for formula validity with respect to classical logic, yet has a quadratic-time decision procedure. We present a sequent-based proof system for quantified orthologic, which we prove sound and complete for the class of all complete ortholattices. We show that orthologic does not admit quantifier elimination in general. Despite that, we show that interpolants always exist in orthologic. We give an algorithm to compute interpolants efficiently. We expect our result to be useful to quickly establish unreachability as a component of verification algorithms.
\end{abstract}

\section{Introduction}

Interpolation-based techniques are important in hardware and software model checking \cite{DBLP:conf/cav/McMillan03,McMillan2018, mcmillanInterpolantsSymbolicModel2007,mcmillanSolvingConstrainedHorn2013,mcmillanInterpolationSATBasedModel2003, henzingerAbstractionsProofs2004,hojjatELDARICAHornSolver2018, kroeningInterpolationBasedSoftwareVerification2011, RuemmerETAL13DisjunctiveInterpolantsHornClauseVerification}. The interpolation theorem for classical propositional logic states that, for two formulas $A$ and $B$ such that $A \implies B$ is valid, there exists a formula $I$, with free variables among only those common to both $A$ and $B$, such that both $A \implies I$ and $I \implies B$ are valid. All known algorithms for propositional logic have worst-case exponential size of proofs they construct, which is not surprising given that the validity problem is coNP-hard. Interpolation algorithms efficiently construct interpolants from such exponentially-sized proofs \cite{pudlakLengthsProofs1998}, which makes the overall process exponential.
It is therefore interesting to explore whether there are logical systems for which proof search and interpolation are polynomial-time in the size of the input formulas.

Orthologic is a relaxation of classical logic, corresponding to the algebraic structure of ortholattices, where the law of distributivity does not necessarily hold, but where the weaker absorption law (V9) does (\autoref{tab:algebraiclaws}). In contrast to classical and intuitionistic logic, where the problem of deciding the validity of a formula is, respectively, coNP-complete and PSPACE-complete, there is a \emph{quadratic-time} algorithm to decide validity in orthologic \cite{guilloudOrthologicAxioms2024,guilloudFormulaNormalizationsVerification2023}. Orthologic was first studied as a candidate for quantum logic, where distributivity fails \cite{birkhoffLogicQuantumMechanics1936,bellOrthologicForcingManifestation1983}. Due to its advantageous computational properties, orthologic has recently been suggested as a tool to reason about proofs and programs in formal verification in a way that is sound, efficient and predictable \cite{guilloudFormulaNormalizationsVerification2023,guilloudLISAModernProof2023}, even if incomplete. 

As a step towards enabling the use of state-of-the-art model checking techniques backed by orthologic, this paper studies interpolation, as well as properties of quantifiers in orthologic.
The quantifier elimination property would immediately lead to the existence of interpolants. We show, however, that quantified orthologic does \emph{not} admit quantifier elimination. To do so, we define semantics of quantified orthologic (QOL) using complete ortholattices. Furthermore, we present a natural sequent calculus proof system for QOL that we show to be sound and complete with respect to this semantics. We then consider the question of interpolation. We show that a refutation-based notion of interpolation fails. However, a natural notion of interpolants based on the lattice ordering of formulas yields interpolants in orthologic. Namely, if $A \leq B$ is provable, then there exists an interpolant $I$ such that $A \leq I$ and $I \leq B$, where $\leq$ corresponds to implication. Moreover, these interpolants can be computed efficiently from a proof of $A \leq B$. We expect that this notion of interpolation can be used in future verification algorithms.
\begin{table}[bth]
    \centering
    \begin{tabular}{r c @{\hskip 2em} | @{\hskip 2em} r c}
         V1: & $x \lor y = y \lor x$  & V1': & $x \land y = y \land x$ \\
         V2: & $x \lor ( y \lor z) = (x \lor y) \lor z$  & V2': & $x \land ( y \land z) = (x \land y) \land z$ \\
         V3: & $x \lor x = x$  & V3': & $x \land x = x$ \\
         V4: & $x \lor 1 = 1$  & V4': & $x \land 0 = 0$ \\
         V5: & $x \lor 0 = x$  & V5': & $x \land 1 = x$ \\
         V6: & $\neg \neg x = x$\\
         V7: & $x \lor \neg x = 1$  & V7': & $x \land \neg x = 0$ \\
         V8: & $\neg (x \lor y) = \neg x \land \neg y$  & V8': &  $\neg (x \land y) = \neg x \lor \neg y$ \\
         V9: & $x \lor (x \land y) = x$ & V9': & $x \land (x \lor y) = x$   
         \vspace{1em}
    \end{tabular}
    
    \caption{Axioms of orthologic, a generalization of classical logic corresponding to the algebraic variety of ortholattices. As lattices, ortholattices admit a partial order $\leq_\OL$ defined by 
$a \leq_\OL b$ iff $a\land b = a$ or, equivalently, $a \lor b = b$.
    \label{tab:algebraiclaws}}
\end{table}

In some cases, ortholattices may be not only a relaxation of propositional logic but a direct intended interpretation of formulas. Indeed, lattices already play a crucial role in abstract interpretation \cite{cousotAbstractInterpretationUnified1977,nielsonPrinciplesProgramAnalysis1999} and have been adopted by the Flix programming language \cite{madsenDatalogFlixDeclarative2016a}. 
Furthermore, De Morgan bi-semilattices and lattices (generalizations of ortholattices where law V6 of \autoref{tab:algebraiclaws} does not necessarily hold) have been used to model multivalued logics with undefined states \cite{brzozowskiMorganBisemilattices2000,brunsModelCheckingMultivalued2004}. Lattice automata \cite{DBLP:conf/vmcai/KupfermanL07} map final automaton states to elements of a finite distributive De Morgan lattice, which admits a notion of complement, but, in contrast to ortholattices, need not satisfy V7 or V7' of~\autoref{tab:algebraiclaws} (the chain $0 \le 1/2 \le 1$ is a finite distributive de Morgan lattice but not an ortholattice).

Proof-theoretic properties of propositional orthologic are presented in \cite{guilloudOrthologicAxioms2024}, but without discussion of interpolation and without the treatment of quantifiers as lattice operators. These topics are the subject of the present paper.

\clearpage
\noindent
{\bf Contributions.}\ 
We make the following contributions:
\begin{enumerate}
    \item We define quantified orthologic, in the spirit of QBF, presenting its semantics in terms of validity in all complete ortholattices. We present a proof system for quantified orthologic, which extends an existing polynomial-time proof system for quantifier-free orthologic \cite{schultemontingCutEliminationWord1981,guilloudOrthologicAxioms2024} with rules for quantifier introduction and elimination. We show soundness and completeness of our proof system.
    
    \item We show that quantified orthologic does not admit quantifier elimination. Consequently, quantifiers increase the class of definable relationships between ortholattice elements.
    This also makes the existence of interpolants a more subtle question than in classical propositional logic, where quantifier elimination alone guarantees that quantifier-free interpolants exist.

    \item We consider a refutation-based interpolation property, which reduces to the usual one in classical logic. We show that orthologic does \emph{not} satisfy this variant of interpolation.

    \item We consider another notion of interpolation, one which is natural in any lattice-based logic. In the language of ortholattices (\autoref{tab:posetlaws}), given two formulas $A$ and $B$ such that $A \leq B$, an interpolant $I$ is a formula such that $A \leq I$, $I \leq B$, and $\freevars{I} \subseteq \freevars{A} \cap \freevars{B}$. While it is known \cite{miyazakiSuperAmalgamationPropertyVariety1999} that orthologic admits such interpolants, we show using the sequent calculus proof system for OL that (a generalization of) such interpolants can always be computed efficiently. Specifically, we present an algorithm to compute interpolants from proofs of sequents in time linear in the size of the proof (where finding proofs in orthologic is worst-case quadratic time in the size of the inequality). 
\end{enumerate}
The final result yields an end-to-end polynomial-time algorithm that first finds a proof and then computes an interpolant $I$, where validity of both the input $A \implies B$ and the result $A \implies I$, $I \implies B$ is with respect to OL axioms.

\begin{table}[bth]
    \centering
    \begin{tabular}{r c @{\hskip 2em} | @{\hskip 2em} r c}
         P1: & $x \leq x$\\
         P2: & $x\leq y \hspace{0.5em} \& \hspace{0.5em} y \leq z \implies x \leq y$\\
         P3: & $0 \leq x$ & P3': & $x \leq 1$\\
         P4: & $x \land y \leq x$ & P5': & $x \leq x \lor y$\\
         P5: & $x \land y \leq y$ & P6': & $y \leq x \lor y$\\
        
         P6: & $x \leq y \hspace{0.5em} \& \hspace{0.5em} x \leq z \implies x \leq y\land z$ & 
            P6': & $x \leq z \hspace{0.5em} \& \hspace{0.5em} y \leq z \implies x\lor y \leq z$\\
         P7: & $x \leq \neg \neg x$ & P7': & $\neg \neg x \leq x$\\
         P8: & $x \leq y \implies \neg y \leq \neg x$\\
         P9: & $x \land \neg x \leq 0$ & P9': & $1 \leq x \lor \neg x$\\
         \vspace{0.5em}
    \end{tabular}
    \caption{Axiomatization of ortholattices in the signature $(S, \leq, \land, \lor, 0, 1, \neg)$ as partially ordered sets. $\&$ denotes conjunction between atomic formulas of axioms, to differentiate it from the term-level lattice operation $\land$.
    \label{tab:posetlaws}}
\end{table}

\noindent
{\bf Preliminaries.}
We follow the definitions and notation of \cite{guilloudOrthologicAxioms2024}.
An ortholattice is an algebraic variety with language $(\land, \lor, \neg, 0,1)$ and axioms in \autoref{tab:algebraiclaws}. As lattices, ortholattices can be described as a partially ordered set whose order relation, noted $\leq_\OL$, is defined by $a \leq_\OL b$ iff $a\land b = a$ or equivalently $a \lor b = b$ \cite{meinanderSolutionUniformWord2010,kalmbachOrthomodularLattices1983}. In both Boolean and Heyting algebras, this order relation corresponds to the usual logical implication. The (equivalent) axiomatization of ortholattices as a poset can be found in \autoref{tab:posetlaws}.
We denote by $\term{}{\OL}$ the set of terms built as trees with nodes labelled by either by
$(\land, \lor, \neg, 0, 1)$ or by symbols in a fixed, countably infinite set of variable symbols 
$\Var = \lbrace x, y, z, ... \rbrace$. This corresponds precisely to the set of propositional formulas.
Note that since $\land$ and $\lor$ are commutative, children of a node are described for simplicity as an unordered set. In particular, $x \land y$ and $y \land x$ denote the same term. Since $0$ can always be represented as $x \land \neg x$, we sometimes omit it from case analysis for brevity, and similarly for $1$.

\section{Quantified Orthologic: Syntax, Semantics, and a Complete Proof System}
We consider the extension of propositional orthologic to quantified orthologic, noted QOL, the analogue of QBF \cite{cookQuantifiedPropositionalCalculus2005} for classical logic, or of System F \cite{girardProofsTypes1989} for intuitionistic logic.
To do so, we extend the proof system of \cite{guilloudOrthologicAxioms2024} by adding axiomatization of an existential quantifier ($\bigvee$) and a universal quantifier ($\bigwedge$). The deduction rules of QOL are in \autoref{fig:proofSystem}. It is folklore that the sequent calculus LK \cite{gentzenUntersuchungenUberLogische1935} with arbitrarily many formulas on both sides corresponds to classical logic, while, if we restrict the right sides of sequents to only contain at most one formula, we obtain intuitionistic logic \cite[section 7.1]{sorensenLecturesCurryHowardIsomorphism2010}. Orthologic exhibits a different natural restriction: sequents can only contain at most two formulas in both sides of the sequent combined.
For this reason, it is convenient to represent sequents by decorating the formulas with superscript ${}^L$ or ${}^R$, depending on whether they appear on the left or right side.
\begin{defin}[From \cite{guilloudOrthologicAxioms2024}]
    If $\phi$ is a formula, we call $\phi^L$ and $\phi^R$ annotated formulas.
    A \emph{sequent} is a set  of at most two annotated formulas. We use uppercase Greek letters (e.g. $\Gamma$ and $\Delta$) to represent sets that are either empty or contain exactly one annotated formula ($|\Gamma| \le 1, |\Delta| \le 1$).
\end{defin}
Given formulas $\phi$ and $\psi$, we thus write $\phi^L, \phi^R$ for a sequent often denoted $\phi \vdash \psi$.

Our use of quantifiers in this paper (quantified orthologic) is different from considering the first-order theory of ortholattices. In particular, the semantic of an existential quantifier ($\bigwedge x.\; t)$ in QOL corresponds to the least upper bound of a (possibly infinite) family of \emph{lattice} elements given by values of term $t$. In contrast, when considering a classical first-order theory of ortholattices, we would build an atomic formula such as $t_1 \le t_2$, obtaining a definite truth or falsehood in the metatheory, and only then apply quantifiers to build formulas such as $\exists x. (t_1 \le t_2)$. Such difference also exists in the case of Boolean algebras \cite{DBLP:journals/tcs/Kozen80}.

\begin{figure}[ht]
\begin{framed}
\center
    \begin{subfigure}{0.8\textwidth}
        \begin{tabular}{c c}
        \ \\
            \multicolumn{2}{c}{            
                \AxiomC{}
                \RightLabel{\text { Hyp}}
                \UnaryInfC{$\phi^L, \phi^R$}
                \DisplayProof
            }\\[3ex]

            \multicolumn{2}{c}{
                \AxiomC{$\Gamma, \psi^R$}
                \AxiomC{$\psi^L, \Delta$}
                \RightLabel{\text{ Cut}}
                \BinaryInfC{$\Gamma, \Delta$}
                \DisplayProof
            }\\[3ex]

            \multicolumn{2}{c}{
                \AxiomC{$\Gamma$}
                \RightLabel{\text { Weaken}}
                \UnaryInfC{$\Gamma, \Delta$}
                \DisplayProof
            }\\[3ex]

            \AxiomC{$\Gamma, \phi^L$}
            \RightLabel{\text { LeftAnd}}
            \UnaryInfC{$\Gamma, (\phi \land \psi)^L$}
            \DisplayProof &
            \AxiomC{$\Gamma, \phi^R$}
            \AxiomC{$\Gamma, \psi^R$}
            \RightLabel{\text{ RightAnd}}
            \BinaryInfC{$\Gamma, (\phi \land \psi)^R$}
            \DisplayProof
            \\[3ex]

            \AxiomC{$\Gamma, \phi^L$}
            \AxiomC{$\Gamma, \psi^L$}
            \RightLabel{\text{ LeftOr}}
            \BinaryInfC{$\Gamma, (\phi \lor \psi)^L$}
            \DisplayProof &
            \AxiomC{$\Gamma, \phi^R$}
            \RightLabel{\text{ RightOr}}
            \UnaryInfC{$\Gamma, (\phi \lor \psi)^R$}
            \DisplayProof
            \\[3ex]

            \AxiomC{$\Gamma, \phi^R$}
            \RightLabel{\text { LeftNot}}
            \UnaryInfC{$\Gamma, (\neg \phi)^L$}
            \DisplayProof &
            \AxiomC{$\Gamma, \phi^L$}
            \RightLabel{\text{ RightNot}}
            \UnaryInfC{$\Gamma, (\neg \phi)^R$}
            \DisplayProof \\[3ex]
        \end{tabular}
        \caption{Deduction rules of propositional Orthologic.}
    \end{subfigure}
\\[2em]
    \begin{subfigure}{0.8\textwidth}
        \begin{tabular}{c c}
            \ \\
            \AxiomC{$\Gamma, \phi[x:=\gamma]^L$}
            \RightLabel{\text { LeftForall}}
            \UnaryInfC{$\Gamma, (\bigwedge x. \phi)^L$}
            \DisplayProof &
            \AxiomC{$\Gamma, \phi[x:=x']^R$}
            \RightLabel{\text{ \shortstack{RightForall \\ {\scriptsize($x'$ not free in $\Gamma$)}} }}
            \UnaryInfC{$\Gamma, (\bigwedge x. \phi)^R$}
            \DisplayProof \\[3ex]

            \AxiomC{$\Gamma, \phi[x:=x']^L$}
            \RightLabel{\text { \shortstack{LeftExists \\ {\scriptsize($x'$ not free in $\Gamma$)}}}}
            \UnaryInfC{$\Gamma, (\bigvee x. \phi)^L$}
            \DisplayProof &
            \AxiomC{$\Gamma, \phi[x:=\gamma]^R$}
            \RightLabel{\text{ RightExists}}
            \UnaryInfC{$\Gamma, (\bigvee x. \phi)^R$}
            \DisplayProof \\[3ex]
\ \\
        \end{tabular}
        \caption{Deduction rules of Quantified Orthologic. }
\end{subfigure}

        \caption{Deduction rules of Orthologic. Each holds for arbitrary $\Gamma$, $\Delta$, $\phi$, $\psi$. Sets $\Gamma$ and $\Delta$ are either empty or contain a single annotated formula.}
\label{fig:proofSystem}

\end{framed}
\end{figure}

\subsection{Complete Ortholattices}
To model quantified Orthologic, we restrict ortholattices to complete ones.

\begin{defin}[Complete Ortholattice]
    An ortholattice $\mathcal O = (O, \sqsubseteq, \sqcup, \sqcap, -)$ is \textit{complete} if and only if for any possibly infinite set of elements $X \subseteq O$, there exist two elements noted $\bigsqcup X$ and $\bigsqcap X$ which are respectively the \textit{lowest upper bound} and \textit{greatest lower bound} of elements of $X$, with respect to $\sqsubseteq$: 
    $$
    \forall x \in X.\ \ x \sqsubseteq \bigsqcup X {\ \text{and}\ } \bigsqcap X \sqsubseteq x~,
    $$
    and, for all $y \in O$:
    $$
    (\forall x \in X. x \sqsubseteq y) \implies (\bigsqcup X \sqsubseteq y)
    $$
    $$
    (\forall x \in X. y \sqsubseteq x) \implies (y \sqsubseteq \bigsqcap X)
    $$
\end{defin}
This definition coincides with the usual definition in complete lattices. Note that, in particular, all finite ortholattices are complete with bounds computed by iterating the binary operators $\sqcup$ and $\sqcap$.

\begin{defin}
    
$\term{}{\QOL}$ denotes the set of quantified orthologic formulas, i.e.
$\term{}{\OL} \subset \term{}{\QOL}$ and for any $x \in \Var$ and $\phi \in \term{}{\QOL}$, $$\bigwedge x. \phi \in \term{}{\QOL}  \text{\ \ and \ \ }  \bigvee x. \phi \in \term{}{\QOL}.$$
Note that $\term{}{\QOL} = \term{}{\QBF}$, the set of quantified Boolean formulas. For two formulas $\phi, \psi \in \term{}{\QOL}$ and a variable $x$, let $\phi[x:=\psi]$ denotes the usual capture-avoiding substitution of $x$ by $\psi$ inside $\phi$. 
\end{defin}
We assume a representation of quantified formulas where alpha-equivalent terms are equal, so that capture-avoiding substitution is well-defined. It is easy to check that any construction in this paper (and in particular, provability) is consistent across alpha-equivalent formulas.

\begin{defin}[Models and Interpretation]\label{defin:models}
    A \textit{model} for QOL is a complete ortholattice $\mathcal O = (O, \sqsubseteq, \sqcup, \sqcap, -)$ and an assignment $\sigma: \Var \rightarrow O$. The interpretation of a formula $\phi$ with respect to an assignment $\sigma$ is defined recursively as usual:
    \[\arraycolsep=4pt\renewcommand{\arraystretch}{1.2}
    \begin{array}{r c l}
        \interp{x}{\sigma}                 & := & \sigma(x)\\
        \interp{\phi \land \psi}{\sigma}   & := & \interp{\phi}{\sigma} \sqcap \interp{\psi}{\sigma}\\
        \interp{\phi \lor \psi}{\sigma}    & := & \interp{\phi}{\sigma} \sqcup \interp{\psi}{\sigma}\\
        \interp{\neg \phi}{\sigma}         & := & -\interp{\phi}{\sigma}\\
        \interp{\bigvee x. \phi}{\sigma}   & := & \bigsqcup \lbrace \interp{\phi}{\sigma[x:=e]} \mid e \in O\rbrace\\
        \interp{\bigwedge x. \phi}{\sigma} & := & \bigsqcap \lbrace \interp{\phi}{\sigma[x:=e]} \mid e \in O\rbrace\\
    \end{array}
    \]
    where $\sigma[x:=e]$ denotes the assignment $\sigma$ with its value at $x$ changed to $e$ and all other values unchanged.

    The interpretation of a sequent is defined in the following way, as in \cite{guilloudOrthologicAxioms2024}:
    \[\arraycolsep=4pt\renewcommand{\arraystretch}{1.2}
    \begin{array}{r c l}
        \interp{\phi^L, \psi^R}{\sigma}     & := & \interp{\phi}{\sigma} \sqsubseteq \interp{\psi}{\sigma}  \\ 
        \interp{\phi^L, \psi^L}{\sigma}     & := & \interp{\phi}{\sigma} \sqsubseteq -\interp{\psi}{\sigma} \\ 
        \interp{\phi^R, \psi^R}{\sigma}     & := & -\interp{\phi}{\sigma} \sqsubseteq \interp{\psi}{\sigma} \\ 
        \interp{\phi^L}{\sigma}             & := & \interp{\phi}{\sigma} \sqsubseteq 0_\mathcal{O} \\ 
        \interp{\phi^R}{\sigma}             & := & 1_\mathcal{O}  \sqsubseteq \interp{\phi}{\sigma} \\ 
        \interp{\emptyset}{\sigma}          & := & 1_\mathcal{O}  \sqsubseteq 0_\mathcal{O} 
    \end{array}
    \]
\end{defin}

\begin{defin}[Entailment]
    If the sequent $\Gamma, \Delta$ is provable, we write $\vdash \Gamma, \Delta$. 
    If for every complete ortholattice $\mathcal O$ and assignment $\sigma: \Var \rightarrow O$, $ \interp{\Gamma, \Delta}{\sigma}$ is true, we write $\vDash \Gamma, \Delta$.
\end{defin}
By slight abuse of notation, we sometimes write, e.g.,
$\phi \vdash \psi$ in place of $\vdash \phi^L, \psi^R$ to help readability.

\newcommand{\equiol}{\dashvdash}
\begin{defin}
Given formulas $\phi$ and $\psi$, let $\phi \equiol \psi$ denote the fact that both $\phi \vdash \psi$ and $\psi \vdash \phi$ are provable.
\end{defin}

\noindent
We show soundness and completeness of QOL with respect to the class of all complete ortholattices. Soundness is easy and direct, completeness less so.
\subsection{Soundness}
\begin{lemma}[Soundness]\label{thm:soundness}
    For every sequent $S$, if $\vdash S$ then $\vDash S$.
\end{lemma}
\begin{proof}
    We simply verify that every deduction rule of \autoref{fig:proofSystem} preserves truth of interpretation in any model. We show the case of LeftAnd as an example, as well as LeftForall and LeftExists. Other cases are easy or analogous.

    Fix an arbitrary ortholattice $\mathcal O = (O, \sqsubseteq, \sqcup, \sqcap, -)$.
    
    \noindent\textbf{LeftAnd:}
    For any assignment $\sigma : \Var \rightarrow O$, the interpretation of the conclusion of a LeftAnd rule is
    $$
    \interp{\Gamma, (\phi \land \psi)^L}{\sigma}
    $$
    $\Gamma$ can be empty, a left formula or a right formula. If it is empty then 
    $$
    \interp{\Gamma, (\phi \land \psi)^L}{\sigma} \iff \interp{0^R, (\phi \land \psi)^L}{\sigma}
    $$
    If $\Gamma = \gamma^L$, then we have
    $$
    \interp{\Gamma, (\phi \land \psi)^L}{\sigma} \iff \interp{(\neg \gamma)^R, (\phi \land \psi)^L}{\sigma}.
    $$
    So without loss of generality we can assume $\Gamma = \gamma^R$ to be a right formula.
    Now,
    \begin{align*}
        \interp{\gamma^R, (\phi \land \psi)^L}{\sigma} & \iff\\
        \interp{\phi \land \psi}{\sigma} \sqsubseteq\interp{\gamma}{\sigma} & \iff \\
        \interp{\phi}{\sigma} \sqcap \interp{\psi}{\sigma} \sqsubseteq\interp{\gamma}{\sigma} & 
    \end{align*}
    But using the premise of the LeftAnd rule and the induction hypothesis, we know $\interp{\gamma^R, \phi^L}{\sigma}$ holds true. Hence,
    \begin{align*}
        \interp{\gamma^R, \phi^L}{\sigma} & \iff \\
        \interp{\phi}{\sigma} \sqsubseteq\interp{\gamma}{\sigma} & \implies \\
        \interp{\phi}{\sigma} \sqcap \interp{\psi}{\sigma} \sqsubseteq\interp{\gamma}{\sigma} & 
    \end{align*}
    where the implication holds by the laws of ortholattices (\autoref{tab:posetlaws}).

    \noindent\textbf{LeftForall:} 
    For any assignment $\sigma : \Var \rightarrow O$,
    $$
    \interp{\Gamma, (\bigwedge x. \phi)^L}{\sigma}
    $$
    where we can again assume $\Gamma = \gamma^R$ without loss of generality. Then
    \begin{align*}
        \interp{\gamma^R, (\bigwedge x. \phi)^L}{\sigma} & = \\
        \interp{\bigwedge x. \phi}{\sigma} \sqsubseteq\interp{\gamma}{\sigma} & = \\
        \bigsqcap \lbrace \interp{\phi}{\sigma[x:=e]} \mid e \in O \rbrace  \sqsubseteq\interp{\gamma}{\sigma} & 
    \end{align*}
    Again, by hypothesis, there exists a formula $\psi$ such that $\interp{\gamma^R, \phi[x:=\psi]^L}{\sigma}$ holds true. Finally,
    \begin{align*}
        \interp{\gamma^R, \phi[x:=\psi]^L}{\sigma} & \iff \\
        \interp{\phi[x:=\psi]}{\sigma} \sqsubseteq\interp{\gamma}{\sigma} & \iff \\
        \interp{\phi}{ \sigma[x:=\interp{\psi}{\sigma}] } \sqsubseteq\interp{\gamma}{\sigma} & \implies \\
        \bigsqcap \lbrace \interp{\phi}{\sigma[x:=e]} \mid e \in O \rbrace  \sqsubseteq\interp{\gamma}{\sigma} & 
    \end{align*}
    where the last implication holds by definition of $\bigsqcap$.
    
    \noindent\textbf{LeftExists:} 
    For any assignment $\sigma : \Var \rightarrow O$,
    $$
    \interp{\Gamma, (\bigvee x. \phi)^L}{\sigma}
    $$
    where we assume one last time without loss of generality that $\Gamma = \gamma^R$. Then
    \begin{align*}
        \interp{\gamma^R, (\bigvee x. \phi)^L}{\sigma} & \iff \\
        \interp{\bigvee x. \phi}{\sigma} \sqsubseteq\interp{\gamma}{\sigma} & \iff \\
        \bigsqcup \lbrace \interp{\phi}{\sigma[x:=e]} \mid e \in O \rbrace  \sqsubseteq\interp{\gamma}{\sigma} & 
    \end{align*}
    By hypothesis, $\interp{\gamma^R, \phi^L}{\tau}$ holds for any assignment $\tau$, and in particular for any $\tau$ of the form $\sigma[x:=e]$. Since $x$ does not appear in $\gamma$, $\interp{\gamma}{\sigma[x:=e]}$ = $\interp{\gamma}{\sigma}$. Hence, for any $e \in O$, each of the following line holds true:
    \begin{align*}
        \interp{\gamma^R, \phi^L}{\sigma[x:=e]} & \iff \\
        \interp{\phi}{\sigma[x:=e]} \sqsubseteq\interp{\gamma}{\sigma[x:=e]} & \iff \\
        \interp{\phi}{\sigma[x:=e]} \sqsubseteq\interp{\gamma}{\sigma} & 
    \end{align*}
    By the least upper bound property of $\bigsqcup$, we obtain as desired the truth of:
        $$
        \bigsqcup \lbrace \interp{\phi}{\sigma[x:=e]} \mid e \in O \rbrace  \sqsubseteq\interp{\gamma}{\sigma} 
        $$ 
\end{proof}

\subsection{Completeness}

In classical propositional logic, we can show completeness with respect to the $\lbrace 0,1\rbrace$ Boolean algebra, which is straightforward. In orthologic, however, we do not have completeness with respect to a simple finite structure; we will need to build an infinite complete ortholattices. The construction is distinct but not entirely unlike that of models for predicate orthologic \cite{miyazakiSuperAmalgamationPropertyVariety1999, bellOrthologicForcingManifestation1983}. In particular, MacNeille completion \cite{macneillePartiallyOrderedSets1937} is used to transform the initial incomplete model into a complete one.

\begin{lemma}[Completeness]
    \label{thm:completeness}
    For any sequent S, $\vDash S$ implies $\vdash S$.
\end{lemma}
\begin{proof}
    We prove the contraposition: if the sequent $S$ is not provable, then there exists a complete ortholattice $\mathcal O = (O, \sqsubseteq, \sqcup, \sqcap, -)$ and an assignment $\sigma: \Var \to O$ such that $\interp{S}{\sigma}$ does not hold.  We construct $O$ from the set of syntactic terms of complete ortholattices themselves, similarly to a free algebra (but with quantifiers). Formally, let $O$ be $\term{}{\QOL}\ssimw{\dashvdash}$, i.e. the quotient set of $\term{}{\QOL}$ by the relation $\dashvdash$.
    
    It is immediate that the function symbols $\land, \lor, \neg$ and relation symbol $\vdash$ of $\term{}{\QOL}$ are consistent over the equivalence classes of $O$, allowing us to extend them to $O$:
    \vspace{-1em}
    \begin{align*}
         \eqclass{\phi}{\dashvdash} \sqcap \eqclass{\psi}{\dashvdash} & := \eqclass{\phi \land \psi}{\dashvdash} \\
         \eqclass{\phi}{\dashvdash} \sqcup \eqclass{\psi}{\dashvdash} & := \eqclass{\phi \lor \psi}{\dashvdash} \\  
         - \eqclass{\phi}{\dashvdash}           & := \eqclass{\neg \phi}{\dashvdash}  \\
         \eqclass{\phi}{\dashvdash} \sqsubseteq \eqclass{\psi}{\dashvdash} & := (\phi \vdash \psi) \text{ is provable}
    \end{align*}
    It is also immediate that $\mathcal O = (O, \sqsubseteq, \sqcup, \sqcap, -)$ satisfies all the laws of ortholattices of \autoref{tab:algebraiclaws}.
    However, to interpret a quantified formula into $\mathcal{O}$, we would need $\mathcal O$ to be complete. It might not be complete, but it is ``complete enough'' to define all upper bounds of interest, as the following lemma shows.
    
    \begin{lemma}\label{lem:Ointerp}
        For any $\sigma: \Var \rightarrow O$, let  $\sigma': \Var \rightarrow \term{}{\QOL}$ be such that for any $x$
        \\ $\eqclass{\sigma'(x)}{\dashvdash} = \sigma(x)$.
        Let $\phi[\sigma']$ denote the simultaneous capture-avoiding substitution of variables in the formula $\phi$ with the assignments in $\sigma'$.

        Then, for any $\phi \in \term{}{\QOL}$, $\interp{\phi}{\sigma}$ exists and $\interp{\phi}{\sigma} = \eqclass{\phi[\sigma']}{\dashvdash}$. 
    \end{lemma}
    \begin{proof}
    First note that $\eqclass{\phi[\sigma']}{\dashvdash}$ is well-defined: it does not depend on the specific choice of assignment we make for $\sigma '$.
    Then, the proof works by structural induction on $\phi$. If it is a variable $x$, 
    $$
    \interp{x}{\sigma} = \sigma(x) = \eqclass{\sigma'(x)}{\dashvdash} = \eqclass{x[\sigma']}{\dashvdash}
    $$ 
    by definition. Then, if $\phi = \phi_1 \land \phi_2$, 
    $$\interp{\phi_1 \land \phi_2}{\sigma} = \interp{\phi_1}{\sigma} \sqcap \interp{\phi_2}{\sigma} = \eqclass{\phi_1[\sigma']}{\dashvdash} \sqcap \eqclass{\phi_2[\sigma']}{\dashvdash} = \eqclass{\phi_1[\sigma'] \land \phi_2[\sigma']}{\dashvdash}$$
    where the first equality is the definition of $\interp{\cdot}{}$, the second equality the induction hypothesis and the third equality is the definition of $\sqcap$ in $\mathcal O$. $\lor$ and $\neg$ are similar.
    Consider now the interpretation of a formula $\interp{\bigvee x. \phi}{\sigma}$. 
    Since alpha-equivalence holds in our proof system and in the definition of the least upper bound, we assume to ease notation that $x$ is fresh with respect to $\sigma$, i.e., that we don't need to signal explicitly capture-avoiding substitution. 
    By definition of $\interp{\cdot}{\sigma}$, we should have:
    $$
    \interp{\bigvee x. \phi}{\sigma}  = \bigsqcup \lbrace \interp{\phi}{\sigma[x:=e]} \mid e \in O\rbrace
    $$
    Does the right-hand side always exist in $\mathcal{O}$? We claim that it does, and that it is equal to $\eqclass{(\bigvee x.\phi)[\sigma']}{\dashvdash}$. Mainly, we need to show that it satisfies the two properties of the least upper bound. First, the \textit{upper bound} property:
    $$
    \forall a \in \lbrace \interp{\phi}{\sigma[x:=e]} \mid e \in O\rbrace, \ \  a \sqsubseteq \eqclass{(\bigvee x.\phi)[\sigma']}{\dashvdash}
    $$
    Which is equivalent to $\forall e \in O$,
    \begin{align}
        \interp{\phi}{\sigma[x:=e]} &\sqsubseteq \eqclass{(\bigvee x.\phi)[\sigma']}{\dashvdash} & \iff \\
        \eqclass{\phi[\sigma'_{[x:=e]}]}{\dashvdash} &\sqsubseteq  \eqclass{(\bigvee x.\phi)[\sigma']}{\dashvdash}  & \iff \\
        \phi[\sigma'_{[x:=e]}] &\vdash (\bigvee x.\phi)[\sigma']  \text{ is provable}& \iff \\
        \phi[\sigma'_{[x:=e]}] &\vdash \bigvee x.\phi[\sigma'_{[x:=x]}] \text{ is provable}
    \end{align}
    where (1) is the desired least upper bound property, (2) is equivalent by induction hypothesis and definition of $\sigma'$, (3) by definition of $\sqsubseteq$ in $\mathcal O$ and (4) by definition of substitution. The last statement is indeed provable:
    
    \begin{center}
        \AxiomC{}
        \RightLabel{\ruleHypothesis}
        \UnaryInfC{$\phi[\sigma_{[x:=e]}]^L, \phi[\sigma_{[x:=e]}]^R$}
        \RightLabel{\ruleRightExists}
        \UnaryInfC{$\phi[\sigma_{[x:=e]}]^L, (\bigvee x.\phi[\sigma_{[x:=x]}'])^R$}
        \DisplayProof
    \end{center}
    Secondly, we need to show the \textit{least} upper bound property:
    $$
    \forall a \in O. (\forall e \in O. \interp{\phi}{\sigma[x:=e]} \sqsubseteq a) \implies (\eqclass{(\bigvee x.\phi)[\sigma']}{\dashvdash} \sqsubseteq a)
    $$
    which is equivalent to
    $$
    \forall \psi \in \term{}{\QOL}. (\forall e \in O. \interp{\phi}{\sigma[x:=e]} \sqsubseteq \eqclass{\psi}{\dashvdash}) \implies (\eqclass{(\bigvee x.\phi)[\sigma']}{\dashvdash} \sqsubseteq \eqclass{\psi}{\dashvdash})
    $$
    Fix an arbitrary $\psi$ and assume $\forall e \in O. \interp{\phi}{\sigma[x:=e]} \sqsubseteq \eqclass{\psi}{\dashvdash}$. Consider a variable $x_2$ which does not appear in $\psi$. Then, we have in particular,
    $$
    \interp{\phi}{\sigma[x:=x_2]} \sqsubseteq \eqclass{\psi}{\dashvdash}.
    $$
    Then, 
    \begin{align*}
        \interp{\phi}{\sigma[x:=x_2]} &\sqsubseteq \eqclass{\psi}{\dashvdash}  & \iff \\
        \eqclass{\phi[\sigma'_{[x:=x_2]}]}{\dashvdash} &\sqsubseteq \eqclass{\psi}{\dashvdash}  & \iff \\
        \phi[\sigma'_{[x:=x_2]}] &\vdash \psi \text{ is provable} & \iff \\
    \end{align*}
    Then using a proof of the last line, we can construct:
        \begin{center}
        \AxiomC{$\phi[\sigma'_{[x:=x_2]}]^L, \psi^R$}
        \RightLabel{\ruleLeftExists}
        \UnaryInfC{$(\bigvee x.\phi[\sigma'_{[x:=x]}])^L, \psi^R$}
        \DisplayProof
    \end{center}
    We finally obtain our second property as desired:
    $$
    \eqclass{(\bigvee x.\phi)[\sigma']}{\dashvdash}  \ \sqsubseteq\  \eqclass{\psi}{\dashvdash}
    $$
    To conclude the proof of \autoref{lem:Ointerp}, the case with $\bigwedge$ instead of $\bigvee$ is symmetrical.
    \end{proof}
    
    Hence, our interpretation in $\mathcal O$ is guaranteed to be well-defined. However, $\mathcal{O}$ is not guaranteed to be complete for arbitrary sets of elements, which our definition of a model requires. To obtain a complete ortholattice, we will apply MacNeille completion to $\mathcal O$.
    \begin{defin}[MacNeille Completion, \cite{macneillePartiallyOrderedSets1937}]
        Given a lattice $L$, there exists a smallest complete lattice $L'$ containing $L$ as a sublattice with an embedding $i: L \rightarrow L'$ preserving the least upper bounds and greatest lower bounds of arbitrary (possibly infinite) subsets of $L$. This is the MacNeille completion of $L$.
    \end{defin}
    Hence, there exists a complete lattice $\mathcal O'$ containing $\mathcal O$ as a sublattice and preserving the existing least upper bounds and greatest lower bound. But we also need $\mathcal O'$ to be an ortholattice, containing $\mathcal O$ as a subortholattice. Fortunately, this is true thanks to a theorem of Bruns.
    \begin{lemma}[Theorem 4.2 of \cite{brunsFreeOrtholattices1976}]
        For every ortholattice $\mathcal O$, its MacNeille completion $\mathcal O'$ admits an orthocomplementation which extends the orthocomplementation of $\mathcal O$.
    \end{lemma}
    \begin{corollary}
    \label{cor:orthoMacNeil}
        There exists an injective ortholattice homomorphism $i: \mathcal O \rightarrow \mathcal O'$ such that 
        $$
        \forall a, b \in O. a \leq_{\mathcal O} b \iff i(a) \leq_{\mathcal O'} i(b)
        $$
        and for any $X\subset O$ such that $\bigsqcup X$ (resp. $\bigsqcap X$) exists, and 
        \begin{align*}
        &i(\bigsqcup X) = \bigsqcup(\lbrace i(x) \mid x \in X\rbrace) \\
        &i(\bigsqcap X) = \bigsqcap(\lbrace i(x) \mid x \in X\rbrace).
        \end{align*}
    \end{corollary}

    We can now finish our completeness proof. Define $\sigma: \Var \to O$ by $\sigma(x) = \eqclass{x}{\dashvdash}$. Then by \autoref{lem:Ointerp}, $\interp{\phi}{\sigma} = \eqclass{\phi}{\dashvdash}$.
    Let $\gamma, \delta$ be the two formulas such that $\interp{S}{\sigma} = (\interp{\gamma}{\sigma} \sqsubseteq \interp{\delta}{\sigma})$, according to \autoref{defin:models}. Remember that the sequent $S$ is not provable by assumption, i.e. $\interp{\gamma}{\sigma} \not\sqsubseteq \interp{\delta}{\sigma}$, and hence:
    $$
    \eqclass{\gamma}{\dashvdash} \not\sqsubseteq_{\mathcal O} \eqclass{\delta}{\dashvdash}
    $$
    from which we deduce 
    $$
    i(\eqclass{\gamma}{\dashvdash}) \not\sqsubseteq_{\mathcal O'}  i(\eqclass{\delta}{\dashvdash})
    $$
    in the ortholattice $\mathcal O'$. 
    We now define $\tau: \Var \rightarrow O'$ such that $\tau(x) = i(\sigma(x))$, implying (by induction and \autoref{cor:orthoMacNeil}) that for any $\phi$, 
    $$
    i(\eqclass{\phi}{\dashvdash}) = \interp{\phi}{\tau}
    $$ 
    and therefore, in $\mathcal O'$:
    $$
    \interp{\gamma}{\tau} \not\sqsubseteq  \interp{\delta}{\tau}
    $$

    We have hence built a model with the complete ortholattice $\mathcal O'$ and the assignment $\tau$ in which $ \interp{\gamma}{\tau} \not\sqsubseteq  \interp{\delta}{\tau}$, so $\interp{S}{\tau}$ does not hold, as desired.
    
\end{proof}

\begin{theorem}
\label{thm:soundcomplete}
    QOL is sound and complete for complete ortholattices, i.e. for any sequent $S$:
    $$
    \vdash S \iff \vDash S
    $$
\end{theorem}

\section{No Quantifier Elimination for Orthologic}

\begin{defin}[Quantifier Elimination]
    A quantified propositional logic admits \emph{quantifier elimination} if for any term $Q$ there exists a quantifier-free term $E$ such that $Q \equiol E$.
\end{defin}
\begin{ex}
    QBF, the theory of quantified classical propositional logic, admits quantifier elimination that replaces the quantified proposition $\exists x. F$ with the proposition $F[x:=0] \lor F[x:=1]$. This quantifier elimination approach is sound over Boolean algebras in general, thanks to the distributivity law.
\end{ex}
\begin{ex}
    The theory of quantified intuitionistic propositional logic does not admit quantifier elimination. Whereas provability in quantifier-free intuitionistic propositional logic corresponds closely to inhabitation in simply typed lambda calculus and is PSPACE-complete \cite{urzyczynInhabitationTypedLambdacalculi1997}, the quantified theory corresponds to System F, and is undecidable \cite{dudenhefnerSimplerUndecidabilityProof2019}.
\end{ex}

Note that quantifier elimination provides a solution to the interpolant problem for QBF (and QOL, if it were to admit quantifier elimination). Indeed, consider a provable sequent $A_{(\vec x, \vec y)}\vdash B_{(\vec y, \vec z)}$, and $\vec x, \vec y, \vec z$ the free variables in $A$ and $B$. We ask for an interpolant such that
$$A_{(\vec{ x}, \vec{ y})} \vdash I_{\vec{ y}} \text{ and } I_{\vec{ y}} \vdash B_{(\vec{ y}, \vec{ z})}$$

By quantifier elimination, there exists a quantifier-free formula $I_{\vec y}$ equivalent to $\bigwedge z. B_{(\vec{ y}, \vec{ z})}$. This $I_{\vec y}$ satisfies the interpolant condition.
\begin{center}
\begin{tabular}{c c}
    \AxiomC{$A_{(\vec x, \vec y)} \vdash B_{(\vec y, \vec z)}$}
    \RightLabel{ RightForall}
    \UnaryInfC{$A_{(\vec{ x}, \vec{ y})} \vdash (\bigwedge z.  B_{(\vec{ y}, \vec{ z})})$}
    \DisplayProof
    \vspace{4ex} & \vspace{4ex}
    \AxiomC{}
    \RightLabel{ Hyp}
    \UnaryInfC{$B_{(\vec{ y}, \vec{ z})} \vdash  B_{(\vec{ y}, \vec{ z})}$}
    \RightLabel{ LeftForall}
    \UnaryInfC{$(\bigwedge z. B_{(\vec{ y}, \vec{ z})}) \vdash B_{(\vec{ y}, \vec{ z})}$}
    \DisplayProof
\end{tabular}
\end{center}
    \vspace{-2em}
However, the next theorem will show that QOL does not admit quantifier elimination in general, even though it still admits interpolation.

\begin{thm} \label{thm:noQE}
    QOL does not admit quantifier elimination. In particular, there exists no quantifier free formula $E$ such that 
    $$E \equiol \bigvee x. \big( \neg x \land (y \lor x) \big)$$
\end{thm}
\begin{proof}
    For the sake of contradiction, suppose such an $E$ exists. 
    Let $y, w_1,...,w_n$ be the free variables appearing in $E$. Since $\bigvee x. \neg x \land (y \lor x)$ is constant with respect to $w_1,...,w_n$, $E$ must be as well, and hence we can assume $E$ only uses $y$ as a variable. Moreover, the laws of \OL{} in \autoref{tab:algebraiclaws} imply that any quantifier free formula whose only variable is $y$ is equivalent to one of $0, 1, y$ or $\neg y$. This can easily be shown by induction on the structure of the formula:
    
    \begin{center}
    \begin{tabular}{l l @{\hskip 12pt}@{\hskip 12pt} l l}
         $0 \land 0$ &$= 0$   &   $0 \land 1$ &$= 1$        \\
         $0 \land y$ &$= 0$   &   $0 \land \neg y$ &$= 0$   \\
         $1 \land 1$ &$= 1$   &   $1 \land y$ &$= y$   \\
         $1 \land \neg y$ &$= \neg y$  &   $y \land  y$ &$= y$   \\
         $y \land \neg y$ &$= 0$  &   $\neg y \land  \neg y$ &$= \neg y$   \\
         $\neg 0$ &$= 1$  &   $\neg 1$ &$= 0$   \\
         $\neg \neg y$ &$= y$  &      \\
    \end{tabular}
    \end{center}
    and similarly for disjunction.

    \noindent Now, consider the ortholattices $M_2$ and $M_4$ in \autoref{fig:M4}:
    
    \begin{figure}[ht]
    \centering

        \fbox{\begin{tikzpicture}[
        roundnode/.style={circle, very thick, minimum size=5mm},
        ]
        \node[roundnode]      (main)                              {};
        \node[roundnode]      (l)         [above=0.5cm of main] {$1$};
        \node[roundnode]      (o)         [below=0.5cm of main] {$0$};
        \node[roundnode]        (na)       [right=0.2cm of main] {$\neg a$};
        \node[roundnode]        (a)       [left=0.2cm of main] {$a$};

        \draw[-] (l) -- (a);
        \draw[-] (l) -- (na);
        \draw[-] (a) -- (o);
        \draw[-] (na) -- (o);
        \end{tikzpicture}}
\qquad
                \fbox{\begin{tikzpicture}[
        roundnode/.style={circle, very thick, minimum size=5mm},
        ]
        \node[roundnode]      (main)                              {};
        \node[roundnode]      (l)         [above=0.5cm of main] {$1$};
        \node[roundnode]      (o)         [below=0.5cm of main] {$0$};
        \node[roundnode]        (na)       [left=0.2cm of main] {$\neg a$};
        \node[roundnode]        (a)       [left=0.5cm of na] {$a$};
        \node[roundnode]        (b)       [right=0.2cm of main] {$b$};
        \node[roundnode]        (nb)       [right=0.5cm of b] {$\neg b$};

        \draw[-] (l) -- (a);
        \draw[-] (l) -- (na);
        \draw[-] (l) -- (b);
        \draw[-] (l) -- (nb);
        \draw[-] (a) -- (o);
        \draw[-] (na) -- (o);
        \draw[-] (b) -- (o);
        \draw[-] (nb) -- (o);
        \end{tikzpicture}}
    \caption{The ortholattices $M_2$ and $M_4$.  $M_2$ is distributive, but $M_4$ is not.}
    \label{fig:M4}
\end{figure}
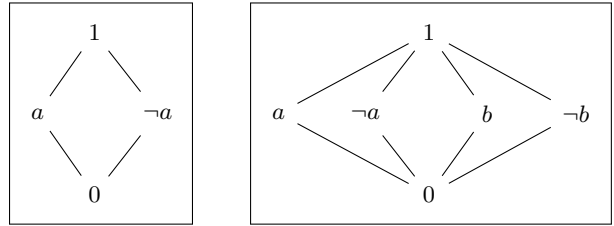
    \noindent We use soundness of orthologic over ortholattices (\autoref{thm:soundness}) so that the formula E (if it exists) needs to be equal to $\bigvee x. \neg x \land (y \lor x) $ in all models. Since the model is finite, it is straightforward to compute in the ortholattice $M_2$ with the assignment $y := a$ that:
    $$\left\llbracket \bigvee x. \neg x \land (y \lor x) \right\rrbracket_{M_2, y:=a} = a$$
    And hence the only compatible formula for $E$ is the atom $y$.

    However, in $M_4$:
    $$\left\llbracket \bigvee x. \neg x \land (y \lor x) \right\rrbracket_{M_4, y:=a} = 1$$
    Hence, any expression for $E$ among $0, 1, y, \neg y$ will fail to satisfy at least one of the two examples, and we conclude that there is no quantifier free formula $E$ that is equivalent to $\bigvee x. \neg x \land (y \lor x)$.
    \qed
\end{proof}

\newcommand{\upclose}[1]{\lceil #1 \rceil}
On one hand, this result shows that we can use quantifiers to define new operators, such as
\[
   \upclose{y} \ \equiv \ \bigvee x. \big( \neg x \land (y \lor x) \big)
\]
while \autoref{thm:noQE} shows that $\upclose{y}$ is not expressible without quantifiers.
On the other hand, this result implies that we cannot use quantifier elimination to compute quantifier-free interpolants; such interpolants require a different approach.

\section{Failure of a Refutation-Based Interpolation}
\label{sec:refutation}
We now consider a notion of interpolation based on orthologic with axioms.
Using axioms makes the assumptions stronger and is a closer approximation of classical propositional logic.

\begin{defin}[Refutation-Based Interpolation] \label{def:refutatInterp}
    Given an inconsistent pair of sequents $A$ and $B$, i.e. there exists a proof of contradiction (the empty sequent) assuming them, a sequent $I$ is said to be a refutation-based interpolant of $(A, B)$ if 
    \begin{itemize}
        \item $I$ can be deduced from $A$ alone,
        \item $I$ and $B$ are inconsistent, and
        \item $\freevars{I} \subseteq \freevars{A} \cap \freevars{B}$.
    \end{itemize}
\end{defin}

We show, by counterexample, that such an interpolant does not exist in general in orthologic.

\begin{thm}
    Given any inconsistent pair \((A, B)\) of sequents, a refutation-based interpolant for it does not necessarily exist in orthologic. In particular, a refutation-based interpolant does not exist for the choice \(A = (z \lor \lnot y) \land (\lnot z \lor \lnot y)^R \) and \(B = (x \land y) \lor (\lnot x \land y)^R\).
\end{thm}

\begin{proof}
For the counterexample, let $A$ be the sequent 
\[
    (z \lor \lnot y) \land (\lnot z \lor \lnot y)^R
\]
and let $B$ be
\[
    (x \land y) \lor (\lnot x \land y)^R~.
\]

We show the proof of inconsistency of $A$ and $B$ in orthologic in \autoref{fig:ab_inconsistency}. For readability and space constraints, the proof is split into four parts, with \autoref{subfig:ab_inconsistency_y} showing a proof of $y^R$ from $B$, \autoref{subfig:ab_inconsistency_z} and \autoref{subfig:ab_inconsistency_negz} showing proofs of $z^R$ and $\neg z^R$ respectively from $y^R$ and $A$, and \autoref{subfig:ab_inconsistency_contra} finally deriving the empty sequent from $z^R$ and $\neg z^R$.
\begin{figure}[ht]
    \centering
    \begin{subfigure}{\textwidth}
        \begin{center}
            \AxiomC{$(x \land y) \lor (\neg x \land y)^R$}
            \AxiomC{}
            \RightLabel{\ruleHypothesis}
            \UnaryInfC{$y^L, y^R$}
            \RightLabel{\ruleLeftAnd}
            \UnaryInfC{$x \land y^L, y^R$}
            \AxiomC{}
            \RightLabel{\ruleHypothesis}
            \UnaryInfC{$y^L, y^R$}
            \RightLabel{\ruleLeftAnd}
            \UnaryInfC{$\neg x \land y^L, y^R$}
            \RightLabel{\ruleLeftOr}
            \BinaryInfC{$(x \land y) \lor (\neg x \land y)^L, y^R$}
            \RightLabel{\ruleCut}
            \BinaryInfC{$y^R$}
            \DisplayProof
        \end{center}
        \caption{Proof of $y^R$ from $B$.}
        \label{subfig:ab_inconsistency_y}
    \end{subfigure}
    \begin{subfigure}{\textwidth}
        \begin{center}
            \AxiomC{$(\neg z \lor \neg y) \land (z \lor \neg y)^R$}
            \AxiomC{}
            \RightLabel{\ruleHypothesis}
            \UnaryInfC{$z^L, z^R$}
            \AxiomC{$y^R$}
            \RightLabel{\ruleLeftNot}
            \UnaryInfC{$\neg y^L$}
            \RightLabel{\ruleWeaken}
            \UnaryInfC{$\neg y^L, z^R$}
            \RightLabel{\ruleLeftOr}
            \BinaryInfC{$z \lor \neg y^L, z^R$}
            \RightLabel{\ruleLeftAnd}
            \UnaryInfC{
            $(\neg z \lor \neg y) \land (z \lor \neg y)^L, z^R$
            }
            \RightLabel{\ruleCut}
            \BinaryInfC{$z^R$}
            \DisplayProof
        \end{center}
        \caption{Proof of $z^R$ from $y^R$ and $A$.}
        \label{subfig:ab_inconsistency_z}
    \end{subfigure}
    \begin{subfigure}{\textwidth}
        \begin{center}
            \AxiomC{$(\neg z \lor \neg y) \land (z \lor \neg y)^R$}
            \AxiomC{}
            \RightLabel{\ruleHypothesis}
            \UnaryInfC{$\neg z^L, \neg z^R$}
            \AxiomC{$y^R$}
            \RightLabel{\ruleLeftNot}
            \UnaryInfC{$\neg y^L$}
            \RightLabel{\ruleWeaken}
            \UnaryInfC{$\neg y^L, \neg z^R$}
            \RightLabel{\ruleLeftOr}
            \BinaryInfC{$\neg z \lor \neg y^L, \neg z^R$}
            \RightLabel{\ruleLeftAnd}
            \UnaryInfC{
            $(\neg z \lor \neg y) \land (z \lor \neg y)^L, \neg z^R$
            }
            \RightLabel{\ruleCut}
            \BinaryInfC{$\neg z^R$}
            \DisplayProof
        \end{center}
        \caption{Proof of $\neg z^R$ from $y^R$ and $A$.}
        \label{subfig:ab_inconsistency_negz}
    \end{subfigure}
    \begin{subfigure}{\textwidth}
        \begin{center}
            \AxiomC{$\neg z^R$}
            \AxiomC{$z^R$}
            \RightLabel{\ruleLeftNot}
            \UnaryInfC{$\neg z^L$}
            \RightLabel{\ruleCut}
            \BinaryInfC{$\emptyset$}
            \DisplayProof
        \end{center}
        \caption{Proof of $\emptyset$ from $z^R$ and $\neg z^R$.}
        \label{subfig:ab_inconsistency_contra}
    \end{subfigure}
    \caption{Proof of inconsistency of $A = (\neg z \lor \neg y) \land (z \lor \neg y)$ and $B = (x \land y) \lor (\neg x \land y)$.}
    \label{fig:ab_inconsistency}
\end{figure}

Given the inconsistent pair of $A$ and $B$, however, we find that the only common variable between $A$ and $B$ is $y$. Sequents built only from the variable $y$ are equivalent to one of $0^R, 1^R, y^R, (\neg y)^R$, none of which is a consequence of $A$ while also being inconsistent with $B$.

Hence, the inconsistent sequent pair $A, B$ as chosen does not admit a refutation-based interpolant according to \autoref{def:refutatInterp}.
\end{proof}
 

\section{Interpolation for Orthologic Formulas}
An arguably more natural definition of interpolation for a lattice-based logic (such as orthologic) using the $\leq$ relation is the following:
\begin{definition}[Implicational interpolation]
    Given two propositional quantifier-free formulas $A$ and $B$ such that $A \leq B$, an interpolant is a formula $I$ such that $\freevars{I} \subseteq \freevars{A} \cap \freevars{B}$ and $A \leq I \leq B$.
\end{definition}
For classical logic, this definition is equivalent to the one of \autoref{sec:refutation}, since $A \leq_\CL B$ if and only if the empty sequent is provable from $(A, \neg B)$. These definitions are however not equivalent in intuitionistic logic and orthologic.

We now prove that the theory of ortholattices admits this form of interpolation by showing a procedure constructing the interpolant inductively from a proof of the sequent \(A^L,B^R\). For induction, we prove a slightly more general statement:
\begin{thm}[Interpolant for orthologic sequents]
\label{thm:deduction_interpolation}
    There exists an algorithm (\lstinline{interpolate} in \autoref{fig:interpolation_algo}) , which,  given a proof of a sequent $\Gamma, \Delta$, computes
    a formula \(I\) called an \emph{interpolant} for the ordered pair $(\Gamma, \Delta)$ 
    such that \(\freevars{I} \subseteq \freevars{\Gamma} \cap \freevars{\Delta}\) and the sequents
    \(\Gamma, I^R\) and \(I^L, \Delta\) are provable. 
    The algorithm has runtime linear in the size of the given proof of $\Gamma, \Delta$.
\end{thm}

Note that interpolants for $(\Gamma, \Delta)$ and for $(\Delta, \Gamma)$ are distinct. In fact, in orthologic, they are negations of each other. Note also that if the sequent $\Gamma, \Delta$ is provable, then by \cite{guilloudOrthologicAxioms2024} there is a proof of it of at most quadratic size. Hence, the interpolation algorithm runs in the worst-case in time quadratic in sizes of $\Gamma$ and $\Delta$.

We have made an executable Scala implementation of the interpolation algorithm alongside orthologic proof
search (as described in \cite{guilloudOrthologicAxioms2024}) open-source on GitHub
\footnote{\url{https://github.com/sankalpgambhir/ol-interpolation}}. 


\newcommand{\backtick}{\ensuremath{{^\backprime}}}

\begin{figure}[hp]
\center
\begin{lstlisting}[language=scala]
def interpolate(
    $\Gamma$: Option[AnnotatedFormula],   
    $\Delta$: Option[AnnotatedFormula],   // the input sequent $\Gamma, \Delta$
    $p$: ProofStep                        // proof of validity of the input sequent
): Formula = 
($\Gamma$, $\Delta$) match
    case (Some($\Pi$), None) => 0
    case (None, Some($\Pi$)) => 1
    case (Some($\Pi$), Some($\Pi$)) => 0 // or 1
    case _ =>
        $p$ match
            case Hypothesis($\phi$) => 
                $\Gamma$ match
                    case Some($\backtick\phi\backtick^L$) => $\phi$
                    case Some($\backtick\phi\backtick^R$) => $\neg\phi$
            case Weaken($\Sigma$, $p'$) =>
                $\Gamma$ match
                    case $\backtick\Sigma\backtick$ => interpolate(None, $\Delta$, $p'$)
                    case _ => interpolate($\Gamma$, None, $p'$)
            case LeftAnd($\phi$, $\psi$, $p'$) =>
                $\Gamma$ match
                    case Some($\backtick\phi\backtick\land \backtick\psi\backtick^L$) => interpolate($\phi^L$, $\Delta$, $p'$)
                    case _ => interpolate($\Gamma$, $\phi^L$, $p'$)
            case RightAnd($\phi$, $\psi$, $p_1$, $p_2$) =>
                $\Gamma$ match
                    case Some($\backtick\phi\backtick \land \backtick\psi\backtick^R$) => 
                        interpolate($\phi^R$, $\Delta$, $p_1$) $\lor$ interpolate($\psi^R$, $\Delta$, $p_2$)
                    case _ => 
                        interpolate($\Delta$, $\phi^R$, $p_1$) $\land$ interpolate($\Delta$, $\psi^R$, $p_2$)
            case LeftOr($\phi$, $\psi$, $p_1$, $p_2$) =>
                $\Gamma$ match
                    case Some($\backtick\phi\backtick \lor \backtick\psi\backtick^L$) => 
                        interpolate($\phi^L$, $\Delta$, $p_1$) $\lor$ interpolate($\psi^L$, $\Delta$, $p_2$)
                    case _ => 
                        interpolate($\Delta$, $\phi^L$, $p_1$) $\land$ interpolate($\Delta$, $\psi^L$, $p_2$)
            case RightOr($\phi$, $\psi$, $p'$) =>
                $\Gamma$ match
                    case Some($\backtick\phi\backtick \lor \backtick\psi\backtick^R$) => interpolate($\phi^R$, $\Delta$, $p'$)
                    case _ => interpolate($\Gamma$, $\phi^R$, $p_1$)
            case LeftNot($\phi$, $p'$) =>
                $\Gamma$ match
                    case Some($\neg\backtick\phi\backtick^L$) => interpolate($\phi^R$, $\Delta$, $p'$)
                    case _ => interpolate($\Gamma$, $\phi^R$, $p'$)
            case RightNot($\phi$, $p'$) =>
                $\Gamma$ match
                    case Some($\neg\backtick\phi\backtick^R$) => interpolate($\phi^L$, $\Delta$, $p'$)
                    case _ => interpolate($\Gamma$, $\phi^L$, $p'$)
\end{lstlisting}

\caption{The algorithm \lstinline|interpolate| to produce an interpolant for any valid sequent, given a partition as an ordered pair and a proof. \lstinline[mathescape=true]|$\backtick\phi\backtick$| in a pattern match is Scala syntax to indicate that $\phi$ is an existing variable to be tested for equality, and not a fresh variable free to be assigned.}
\label{fig:interpolation_algo}
\end{figure}

\begin{proof}
    We show correctness of the algorithm in \autoref{fig:interpolation_algo} with inputs \(\Gamma, \Delta, P\) where $P$ is a proof of the sequent $S = \Gamma, \Delta$. By the cut-elimination theorem of orthologic \cite{guilloudOrthologicAxioms2024,schultemontingCutEliminationWord1981}, we assume that $P$ is cut-free. We show that the result of \lstinline|interpolate|\((\Gamma, \Delta, P)\) is an interpolant for \((\Gamma, \Delta)\). 

    We first deal with the particular case where either $\Gamma$ or $\Delta$ is empty or when $\Gamma = \Delta$, as it will simplify the rest of the proof to assume that they are both non-empty and distinct.
    \begin{itemize}
        \item Suppose $(\Gamma, \Delta) = (\Pi, \emptyset)$. Then $0$ is an interpolant, as both
        $\Pi, 0^R \text{ and } 0^L, \emptyset$
        are provable.
        \item Suppose $(\Gamma, \Delta) = (\emptyset, \Pi)$. Then $1$ is an interpolant, as both
        $\emptyset, 1^R \text{ and } 1^L, \Pi$
        are provable
        \item Suppose $(\Gamma, \Delta) = (\Pi, \Pi)$. Then any formula $\psi$ (in particular 0 and 1) is an interpolant as both
        $\Pi, \psi^R \text{ and } \psi^L, \Pi$
        are provable by weakening.
    \end{itemize}
    
    In all other cases, the algorithm works recursively on the proof tree of \(P\), starting from the
    concluding (root) step. At every step, the algorithm reduces the construction of the interpolant of \(S\) to those of its premises.
    By induction, assume that for a given proof $P$, the algorithm is correct for all proofs of smaller size (and in particular for the premises of $P$) and consider every proof step from \autoref{fig:proofSystem} with which $P$ can be concluded:

    \begin{itemize}
        \item \ruleHypothesis :  suppose the concluding step is
        \begin{gather*}
            \AxiomC{}
            \RightLabel{\ruleHypothesis~,}
            \UnaryInfC{\(\phi^L, \phi^R\)}
            \DisplayProof
        \end{gather*}
        We must have \((\Gamma, \Delta) = (\phi^L, \phi^R)\), or \((\Gamma, \Delta) =
        (\phi^R, \phi^L)\). Assuming the former, consider the
        interpolant \(I = \phi\). We then trivially have proofs of \((\Gamma, I^R) = (\phi^R, \phi^L)\) and \((I^L, \Delta^R) = (\phi^L, \phi^R)\):

                \begin{center}
        \begin{tabular}{c c c}
            \AxiomC{}
            \RightLabel{\ruleHypothesis~,}
            \UnaryInfC{\(\phi^L, \phi^R\)}
            \DisplayProof &
            
            \hspace*{1em} and \hspace*{1em} &
            
            \AxiomC{}
            \RightLabel{\ruleHypothesis~,}
            \UnaryInfC{\(\phi^L, \phi^R\)}
            \DisplayProof
            \\
        \end{tabular}
        \end{center}

        The latter case is symmetrical, with \(I = \neg\phi\).
    


        \item \ruleWeaken :  suppose the final inference is
        \begin{gather*}
            \AxiomC{\(\Pi\)}
            \RightLabel{\ruleWeaken~.}
            \UnaryInfC{\(\Pi, \Sigma\)}
            \DisplayProof
        \end{gather*}
        As before, we must have \((\Gamma, \Delta) = (\Pi, \Sigma)\) or \((\Gamma, \Delta) =
        (\Sigma, \Pi)\). In the former case, consider the interpolant $C$ for
        \((\Pi, \emptyset)\), the premise (in fact $C = 0$ or $C = 1$). By the hypothesis, the sequents
        \begin{gather*}
            \Pi, C^R \text{ , and\hspace{2ex} } C^L, \emptyset
        \end{gather*}

        are provable. Taking $I = C$, and applying \ruleWeaken on the second sequent, we obtain proofs of $\Gamma, I^R$ and $I^L, \Delta$:

        \begin{center}
        \begin{tabular}{c c c}
             \AxiomC{\(\Pi, C^R\)}
            \DisplayProof &
            
            \hspace*{1em} and \hspace*{1em} &
            
            \AxiomC{\(C^L\)}
            \RightLabel{\ruleWeaken}
            \UnaryInfC{\(C^L, \Delta\)}
            \DisplayProof
            \\
        \end{tabular}
        \end{center}

        The case \(\Gamma = \Delta\) is analogous, with \(I = \neg C\).

        \item \ruleLeftAnd :  suppose the final inference is
        \begin{gather*}
            \AxiomC{\(\Pi, \phi^L\)}
            \RightLabel{\ruleLeftAnd~.}
            \UnaryInfC{\(\Pi, \phi \land \psi^L\)}
            \DisplayProof
        \end{gather*}

        We must have \((\Gamma, \Delta) = ((\phi \land \psi)^L, \Pi)\) or
        swapped. In the former case, by the induction hypothesis, consider an
        interpolant \(C\) for \((\phi^L, \Pi)\), such that the sequents
        %
        %
        \begin{center}
            \begin{tabular}[ht]{p{0.2\textwidth} l}
                $\phi^L, C^R$ & $C^L, \Pi$
            \end{tabular}
        \end{center}

        are provable. For \(I = C\) as interpolant, we have proofs of $\Gamma, I^R$ and $I^L, \Delta$:
        \begin{center}
        \begin{tabular}{c c c}
            \AxiomC{\(\phi^L, C^R\)}
            \RightLabel{\ruleLeftAnd}
            \UnaryInfC{\(\phi \land \psi^L, C^R\)}
            \DisplayProof &
            
            \hspace*{1em} and \hspace*{1em} &

            \AxiomC{\(C^L, \Pi\)}
            \DisplayProof~.
            \\
        \end{tabular}
        \end{center}

        Since \(\freevars{\phi} \subseteq \freevars{\phi \land \psi}\), \(I =
        C\) is an interpolant for the conclusion as required. The case where
        \((\Gamma, \Delta) = (\Pi, (\phi \land \psi)^L)\) is analogous.

        \item \ruleRightAnd :  suppose the final inference is
        \begin{gather*}
            \AxiomC{\(\Pi, \phi^R\)}
            \AxiomC{\(\Pi, \psi^R\)}
            \RightLabel{\ruleRightAnd~.}
            \BinaryInfC{\(\Pi, (\phi \land \psi)^R\)}
            \DisplayProof
        \end{gather*}

        We have \((\Gamma, \Delta) = (\Pi, (\phi \land \psi)^R)\), or the other way round. Assume the former. Applying the induction hypothesis twice, we obtain an
        interpolant for each of the premises, \(C_\phi\) and \(C_\psi\), such
        that the sequents            
        
        \begin{center}
            \begin{tabular}[ht]{p{0.2\textwidth} p{0.2\textwidth}} \(\Pi,
                C_\phi^R\) & \(C_\phi^L, \phi^R\) \\
                \(\Pi, C_\psi^R\) & \(C_\psi^L, \psi^R\)
            \end{tabular}
        \end{center}

        are valid. Take \(I = C_\phi \land C_\psi\) as  interpolant. Indeed, its free variables are contained in
        \(\freevars{\Pi} \cap (\freevars{\phi} \cup \freevars{\psi}) =
        \freevars{\Gamma} \cap \freevars{\phi \land \psi}\).

        We then need proofs for $\Gamma, I^R$ and $I^L, \Delta$:
        \begin{gather*}
            \AxiomC{\(\Pi, C_\phi^R\)}
            \AxiomC{\(\Pi, C_\psi^R\)}
            \RightLabel{\ruleRightAnd}
            \BinaryInfC{\(\Pi, (C_\phi \land C_\psi)^R\)}
            \DisplayProof\\\\
            \text{and}\\\\
            \AxiomC{\(C_\phi^L, \phi^R\)}
            \RightLabel{\ruleLeftAnd}
            \UnaryInfC{\(C_\phi \land C_\psi^L, \phi^R\)}
            \AxiomC{\(C_\psi^L, \psi^R\)}
            \RightLabel{\ruleLeftAnd}
            \UnaryInfC{\(C_\phi \land C_\psi^L, \psi^R\)}
            \RightLabel{\ruleRightAnd.}
            \BinaryInfC{\((C_\phi \land C_\psi)^L, (\phi \land \psi)^R\)}
            \DisplayProof
        \end{gather*}
        showing that \(I\) is an interpolant for the pair \((\Gamma, \Delta)\).

        Now in the other case \((\Gamma, \Delta) = ((\phi \land \psi)^R, \Pi)\), the induction hypothesis gives us the following interpolants:
        \begin{center}
            \begin{tabular}[ht]{p{0.2\textwidth} l} \(\phi^R,
                D_\phi^R\) & \(D_\phi^L, \Pi\) \\
                \(\psi^R, D_\psi^R\) & \(D_\psi^L, \Pi\)
            \end{tabular}
        \end{center}
        We can then take $I = (D_\phi \lor D_\psi)$ to obtain proofs of $\Gamma, I^R$ and $I^L, \Delta$:
        \begin{gather*}
            \AxiomC{\(\phi^R, D_\phi^R\)}
            \RightLabel{\ruleRightOr}
            \UnaryInfC{\(\phi^R, (D_\phi \lor D_\psi)^R\)}
            \AxiomC{\(\psi^R, D_\psi^R\)}
            \RightLabel{\ruleRightOr}
            \UnaryInfC{\(\psi^R, (D_\phi \lor D_\psi)^R\)}
            \RightLabel{\ruleRightAnd.}
            \BinaryInfC{\((\phi \land \psi)^R, (D_\phi \lor D_\psi)^R\)}
            \DisplayProof\\\\
            \text{and}\\\\
            \AxiomC{\(D_\phi^L, \Pi\)}
            \AxiomC{\(D_\psi^L, \Pi\)}
            \RightLabel{\ruleLeftOr}
            \BinaryInfC{\((D_\phi \lor D_\psi)^L, \Pi\)}
            \DisplayProof
        \end{gather*}

        Note that we can show by induction that \(D_\phi \lor D_\psi = \neg (C_\phi \land C_\psi)\).
        
        \item \ruleLeftNot :  suppose the final inference is
        \begin{gather*}
            \AxiomC{\(\Pi, \phi^R\)}
            \RightLabel{\ruleLeftNot~.}
            \UnaryInfC{\(\Pi, \neg \phi^L\)}
            \DisplayProof
        \end{gather*}
        We have \((\Gamma, \Delta) = (\Pi, (\neg \phi)^L)\), or the other way round. Assume the former.
        We apply the induction hypothesis as before to obtain an interpolant \(C\) for \((\Pi, \phi^R)\) such that
        \begin{center}
            \begin{tabular}[ht]{p{0.2\textwidth} l}
                $\Pi, C^R$ & $C^L, \phi^R$
            \end{tabular}
        \end{center}
        are valid. \(I = C\) suffices as an interpolant for the concluding
        sequent, with the proofs of $\Gamma, I^R$ and $I^L, \Delta$
                \begin{center}
        \begin{tabular}{c c c}
            \AxiomC{\(\Pi, C^R\)}
            \DisplayProof &
            
            \hspace*{1em} and \hspace*{1em} &
            
            \AxiomC{\(C^L, \phi^R\)}
            \RightLabel{\ruleLeftNot}
            \UnaryInfC{\(C^L, \neg\phi^L\)}
            \DisplayProof~.
            \\
        \end{tabular}
        \end{center}
    \end{itemize}

    The proofs for the remaining proof rules, \ruleLeftOr, \ruleRightOr, and
    \ruleRightNot, are analogous to the cases listed above.
\end{proof}

\begin{corollary}[Interpolation for Ortholattices]
    Ortholattices admit interpolation, i.e., for any pair of formulas \(A, B\) in an 
    ortholattice with \(A \leq B\), there exists a formula \(I\) such that \(A \leq I\)
    and \(I \leq B\), with $\freevars{I} \subseteq \freevars{A} \cap
    \freevars{B}$.
\end{corollary}


\section{Further Related Work}

The best known interpolation result is Craig's interpolation theorem for first order logic \cite{craigThreeUsesHerbrandGentzen1957}, of which interpolation for classical propositional logic is a special case. Interpolation for predicate intuitionistic logic was first shown by \cite{schuetteInterpolationssatzIntuitionistischenPraedikatenlogik1962}. The propositional case was further studied by \cite{lavaletteInterpolationFragmentsIntuitionistic1989}.

Interpolation can be leveraged, among other applications, to solve constrained Horn clauses \cite{mcmillanSolvingConstrainedHorn2013}, for model checking \cite{mcmillanInterpolationSATBasedModel2003,bradleySATBasedModelChecking2011} or for invariant generation \cite{kovacsFindingLoopInvariants2009,mcmillanQuantifiedInvariantGeneration2008}. Interpolants are computed by many existing solvers and provers such as Eldarica \cite{hojjatELDARICAHornSolver2018}, Vampire \cite{hoderInterpolationSymbolElimination2010} and Wolverine \cite{kroeningInterpolationBasedSoftwareVerification2011}.

A sequent calculus proof system for orthologic was first described in \cite{schultemontingCutEliminationWord1981}, with cut elimination. \cite{chajdaImplicationOrthologic2005} studied implication symbols in orthologic. \cite{meinanderSolutionUniformWord2010} showed that orthologic with axioms is decidable. \cite{miyazakiSuperAmalgamationPropertyVariety1999} showed that Orthologic admits the super amalgamation property, which implies that it admits interpolants. They also show that a predicate logic extension to orthologic admits interpolants, although the proof of both theorems are non-constructive and contain no discussion of algorithms or space and time complexity.
Recently, orthologic was used in practice in a proof assistant \cite{guilloudLISAModernProof2023}, for modelling of epistemic logic \cite{hollidayFundamentalNonClassicalLogic2023,hollidayOrthologicEpistemicModals2022} and for normalizing formulas in software verification \cite{guilloudFormulaNormalizationsVerification2023}.


\section{Conclusion}
We showed that quantified orthologic, with a sequent-based proof system, is sound and complete with respect to all complete ortholattices. A soundness and completeness theorem typically allows demonstrating further provability results by using semantic arguments. We then showed that orthologic does not admit, in general, quantifier elimination. If such a procedure existed, it would have also allowed computing strongest and weakest interpolants. We instead presented an efficient algorithm, computing orthologic interpolants for two formulas, given a proof that one formula implies the other. Computing interpolants is a key part of some algorithms in model checking and program verification. Since orthologic has efficient algorithms to decide validity and compute proofs, which are necessary to compute interpolants, we expect that our present results will allow further development of orthologic-based tools and efficient algorithms for model checking and program verification.

{
\bibliographystyle{splncs04}
\raggedright
\bibliography{more.bib,sguilloud.bib,vkuncak.bib}
}

\end{document}